%% file: main.tex
\documentclass[letterpaper,twocolumn,10pt]{article}
\usepackage{usenix2019_v3}

\usepackage{tikz}
\usepackage{amsmath}
\usepackage{booktabs}
\usepackage{multirow}

\begin{document}

\title{Automatic Cross-Replica Sharding of Weight Update in Data-Parallel Training}
\author{
\begin{tabular}{cccccc}
   Yuanzhong Xu  & HyoukJoong Lee & Dehao Chen & Hongjun Choi & Blake Hechtman & Shibo Wang \\
     \multicolumn{6}{c}{Google}
\end{tabular}
}

\maketitle

\begin{abstract}
In data-parallel synchronous training of deep neural networks, different devices (replicas) run the same program with different partitions of the training batch, but weight update computation is repeated on all replicas, because the weights do not have a batch dimension to partition. This can be a bottleneck for performance and scalability in typical language models with large weights, and models with small per-replica batch size which is typical in large-scale training. This paper presents an approach to automatically shard the weight update computation across replicas with efficient communication primitives and data formatting, using static analysis and transformations on the training computation graph. We show this technique achieves substantial speedups on typical image and language models on Cloud TPUs, requiring no change to model code.

This technique helps close the gap between traditionally expensive (ADAM) and cheap (SGD) optimizers, as they will only take a small part of training step time and have similar peak memory usage. It helped us to achieve state-of-the-art training performance in Google's MLPerf 0.6 submission~\cite{mlperf0.6, google-mlperf0.6}.
\end{abstract}




\newcommand{\todo}[1]{{\color{red} TODO: #1}}

\input{introduction}
\input{overview}
\input{analysis}
\input{transform}
\input{comm}
\input{eval}
\input{related}
\input{conclusion}


\bibliography{references}
\bibliographystyle{acm}

\end{document}

%% file: introduction.tex
\section{Introduction}
\label{intro}

With increasing complexity and data size in deep neural networks, it has become a common practice to leverage distributed, heterogeneous computing devices to parallelize the training process. In the recent MLPerf~0.6 results~\cite{mlperf0.6}, multiple submitters have used over 1000 devices to dramatically reduce the training time.

Data parallelism~\cite{krizhevsky12} is the most commonly used synchronous distributed training strategy due to its simplicity and efficiency. Participating devices are called replicas, which run the same training program that contains the entire neural network, but each replica receives a different partition of the training data batch. Replicas compute their local gradients with their own training data, then communicate to get the combined gradients and apply the same update to their copies of the weight variables. The weight update computation is repeated on all replicas, because the weights and gradients do not have a batch dimension to partition. The cost of weight update on each replica stays constant, even if more devices are added to reduce the per-replica batch size. Due to Amdahl's law, weight update can be a significant overhead for training performance and limit scalability for models with large weights (typical in language models), or small per-replica batch size (typical in large-scale training).

It is natural to think about sharding the weight update computation across replicas, instead of having them all execute the full computation. However, naively sharding the weight update could dramatically increase the data formatting and communication overhead across replicas. First, partitioning a tensor efficiently is non-trivial on modern accelerators with tiled memory layouts~\cite{tpu-perf-guide}. Second, because the forward and backward passes are already partitioned along the batch dimension across replicas, they must receive the full weight in the next training step. In addition to general challenges for efficient communication primitives, a complication is that modern optimizers~\cite{adam,momentum} often require a few auxiliary variables for each weight variable, such as moving average and momentum, each of which has the same size as the weight itself. These auxiliary variables also need to be updated along with the weight. Without weight-update sharding, replicas only need to communicate the gradients; with weight-update sharding, replicas need to communicate the weights and the auxiliary variables, so it is critical to reduce this overhead.

In this paper, we show that with static analysis and proper graph transformations, efficient weight-update sharding can be achieved without any change to the model. Static analysis is used for two purposes, correctness and performance. The correctness analysis identifies parts of the computation graph that are repeated in all replicas, which are the candidate operations for sharding. The performance analysis uses the knowledge of control flow in the graph to find the best places for communication, and estimates the profit of sharding for each part of the repeated computation. With the analysis result, the graph can be transformed with sharded operations, and communication primitives can be added in proper places. We also found that the transformation often enables more advanced optimizations due to reduced live ranges of the the full weight tensors.

The efficiency of communication primitives can be highly affected by the way we shard a multi-dimensional weight tensor, as well as the topology of the training cluster. Our graph transformation carefully chooses the sharding format for each tensor so that it can be efficiently sharded and unsharded. We use different sharding strategies for small- and large-scale training: for small-scale training, we prioritize reducing the shard size since the number of replicas is small; for large-scale training, we instead prioritize reducing the latency of communication.

We have implemented this approach in TensorFlow~\cite{tensorflow} with XLA~\cite{xla}, where most of the analysis and transformation passes are in XLA. XLA is a functional representation where side effects only exist in a few operations, which greatly reduces the complexity of analysis and transformation. We have evaluated the performance improvements for several common image and language models on Cloud TPUs~\cite{tpu}.

%% file: overview.tex
\section{Overview}
\label{overview}
Although our approach can be applied to other systems, we use XLA as the basis for static analysis and graph transforms. We briefly go over some key concepts in XLA, and then give an overview of our approach.

\subsection{XLA background}
XLA is an intermediate representation and its optimizing compiler for linear algebra operations targeting multiple backends (e.g., GPU, TPU, and other accelerators). XLA is currently deployed as a compiler for TensorFlow.

An XLA computation is modeled as a data-flow graph (HLO IR), where nodes represent operators and edges represent the inputs and outputs of each operator. The ordering of operators is solely enforced by data dependencies between the operators~\footnote{A token-type data edge is used to order side-effect operators, if needed.}. Most operators are pure, except for those side-effecting operators to interact between host and device or between devices.

Operator shapes in XLA are static and this restriction enables aggressive compiler optimizations such as buffer assignment, tiling, and rematerialization. Those are key optimizations on accelerators in general as accelerators are often designed with vector/tiled compute units and have limited amount of memory. XLA runs a set of target-independent optimizations (e.g., common-subexpression elimination) as well as target-specific optimizations (e.g., layout assignment, fusion). After running all HLO-level optimizations, the backend-compiler component \textit{lowers} each operator to a lower-level, target-specific representation, and eventually generates low-level machine code for the target.  

Here we list the operators involved during the transformation for the weight-update sharding optimization we present in this paper. Please refer to XLA operational semantics document~\cite{xla} for the full list.

\paragraph{Control flow.} XLA represents control flow as special operators that call nested computations. There are two control-flow operators: \texttt{While} and \texttt{Conditional}. \texttt{While} takes a single operand (of shape \texttt{T}) as the initial value of the loop carried variable and repeats executing its \texttt{body} computation (\texttt{T} $\Rightarrow$ \texttt{T}) until its \texttt{condition} computation (\texttt{T} $\Rightarrow$ \texttt{bool}) returns false. The result of the \texttt{While} is the loop variable (shape \texttt{T}) of the last iteration.

\texttt{Conditional} with N branches takes an operand for the branch index (or a boolean predicate for a 2-way branch) and an operand for the arguments of each branch. \texttt{Conditional} also takes a computation for each branch, where the argument shape of each computation must match the shape of the corresponding operand. The result shape must be the same for all branch computations and this is the result shape of the \texttt{Conditional} operator.  

\paragraph{All-reduce.}
\texttt{All-reduce} has semantics to MPI All-reduce~\cite{MPI-2.2}, which reduces a tensor across participating devices based on the provided binary reduction computation. All-reduce can optionally take a subgroup information, so that the reduction is only applied within each subgroup of devices. For example, a subgroup of $\left\{\left\{0,1,2,3\right\}, \left\{5,6,7,8\right\}\right\}$ combines values among devices 0-3 and devices 4-7 separately.

\paragraph{Data formatting operators.}
\texttt{Transpose} and \texttt{Reshape} are operators used to change the logical shape of tensors. \texttt{Transpose} reorders the dimensions of a tensor based on the permutation pattern (e.g., F32[5,3,8] $\Rightarrow$ F32[3,8,5]), and \texttt{Reshape} changes the shape to a new configuration (e.g., F32[5,3,8] $\Rightarrow$ F32[15,8]). It is important to note that data-formatting semantics is applied to the logical shape, which is not necessarily the same as how the data is laid out physically in memory. For example, a \texttt{Transpose} from a F32[5,3] tensor laid out as row-major to a F32[3,5] tensor laid out as column-major does not need any data movement, and the compiler converts it into a \texttt{Bitcast} operator which is effectively a no-op. 

\begin{figure}[ht]
\centering
\includegraphics[width=0.26\textwidth]{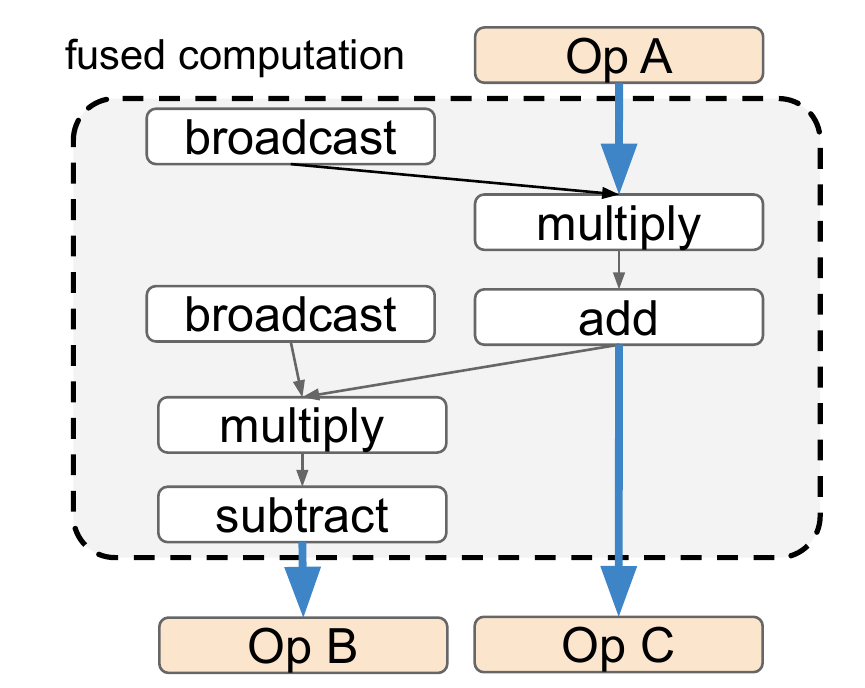}
\caption{\label{f:fusion} A \texttt{Fusion} operator with element-wise operators. Edges in blue represent data transfers from/to the global memory, and all intermediate results are stored in local memory.}
\end{figure}

\paragraph{Fusion operators.}
A \texttt{Fusion} operator represents a group of operators that can be emitted as a unit of computation by the backend of the target device. The fusion optimization pass groups operators that are fusible and replaces them with a fusion operator along with a fusion sub-computation. In the common case, this means that the intermediate results of fused operators are stored in registers or scratchpad memory, without moving data from/to the global memory to save memory bandwidth. Figure~\ref{f:fusion} shows an example of several element-wise operators fused into a single operator.

A more advanced use of \texttt{Fusion} operators would be for the backend compiler to pattern match on the operators within the fusion sub-computation and generate a custom implementation that is semantically equivalent to the original one. Fusion operators used for the weight-update sharding optimization (reduce-scatter and all-gather fusion) correspond to this use case.

\paragraph{Side-effecting operations.}
A small number of operators in XLA are marked as side-effecting and compiler passes need extra care when applying optimizations around the side-effecting operators such that visible side-effects remain the same. Examples are operators used for data transfers between different address spaces, such as \texttt{Infeed} (host to device), \texttt{Outfeed} (device to host), and \texttt{Send/Recv} (device to device).

\subsection{Sharding weight update}
\begin{figure}[ht]
\centering
\includegraphics[width=0.48\textwidth]{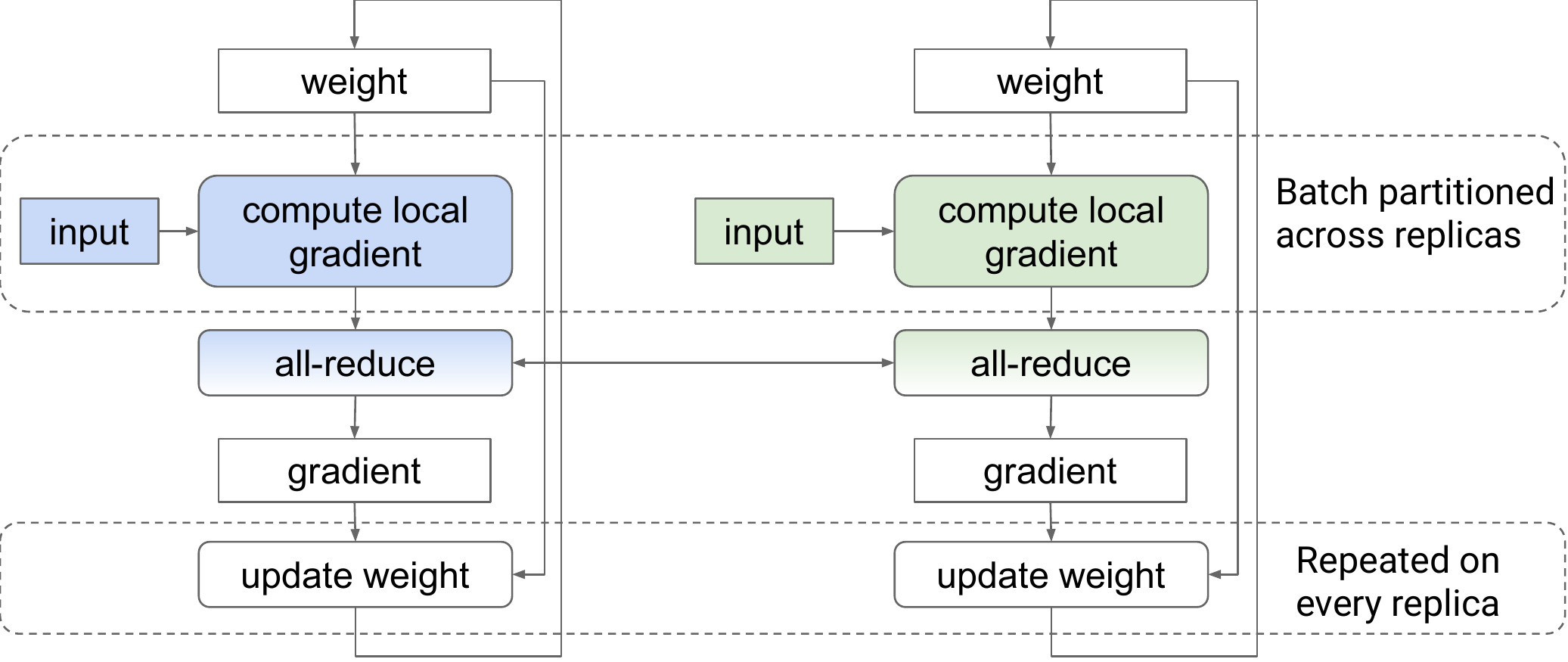}
\caption{\label{f:replication} Synchronous data-parallel training with 2 replicas.}
\end{figure}

Figure~\ref{f:replication} shows a typical synchronous training scenario in data parallelism with two replicas. In every training step, each replica computes its local gradients with its own partition of the training input batch, then all replicas use an all-reduce operator to get the summed gradients. At the end of the training step, all replicas apply the same summed gradients to their copies of the weights, which ensures them to always have the weights in sync as long as they start with the same initial weights.

The training step can spend a non-trivial amount of time in weight update. This is typically true in models with large weights, such as language models like Transformer~\cite{transformer}; in image models like ResNet~\cite{resnet}, although the weight size is usually smaller, when they are trained in large-scale setups with many devices, the per-core batch size is usually set to a small value to avoid excessively large global batch size, making weight update relatively expensive as well. Weight update is memory bound: the compute is mostly simple elementwise operations, but for every weight variable it needs to read the gradient, the original weight and the auxiliary variables, then write back the updated weight and auxiliary variables. In our experiments, Transformer training can spend more than 40\% of the step time in weight update on 1024 TPUv3 chips.

Weight update is not sharded in data parallelism because the weights and gradients do not have a batch dimension to be partitioned. Our goal is to enable sharded weight update across the replicated devices as an optimization, without using more devices.

\paragraph{Sharding with decomposed all-reduce.}
A typical efficient implementation of all-reduce has two phases~\cite{cho19}: reduce-scatter and all-gather. In the {\bf reduce-scatter} phase, replicas exchange data in several rounds on different shards of the data, and at the end,  each replica has one shard of the fully reduced data from all replicas. In the {\bf all-gather} phase, they perform new exchanges to broadcast their own fully reduced shards to all other replicas.

\begin{figure}[ht]
\centering
\includegraphics[width=0.48\textwidth]{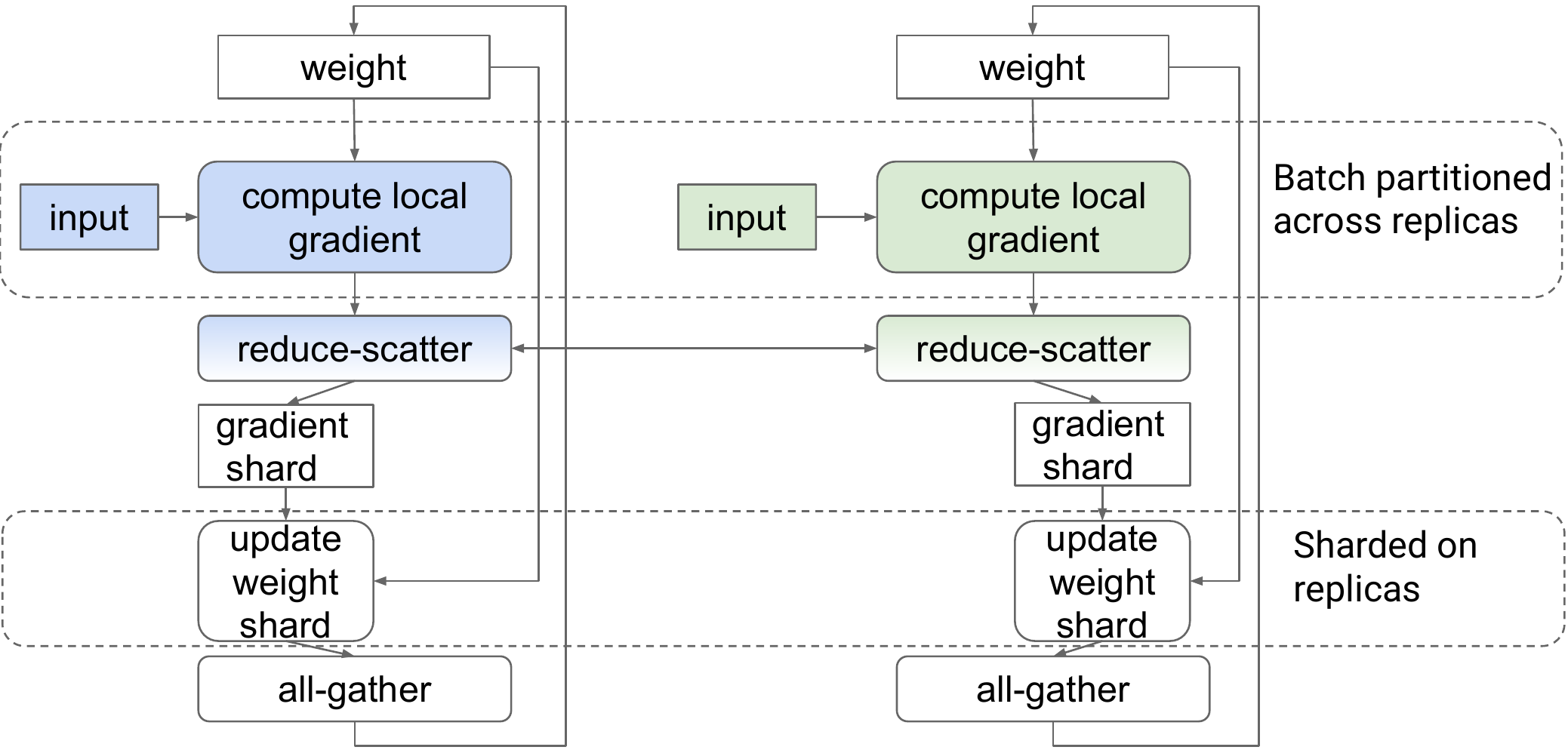}
\caption{\label{f:basic-sharding} Sharding with reduce-scatter an all-gather.}
\end{figure}

As shown in Figure~\ref{f:basic-sharding}, we could use reduce-scatter to produce per-replica shards of the summed gradients, so that each replica can perform weight update on a shard. After that, we could use all-gather to broadcast the updated weight shards to all replicas. The reduce-scatter and all-gather combined should have similar performance as the original all-reduce.

A complication is the use of auxiliary variables in the optimizer. For example, for each weight, the Adam optimizer~\cite{adam} maintains two variables for exponential moving averages of the gradients and squared gradients. These variables are part of the training state and are included in model checkpoints, so typically the updated values are also part of the training step's output. If we do all-gather on every auxiliary variable at the end of each training step, the communication overhead would be too large. However, these variables are only used by the optimizer at the weight-update time, and not needed by the next iteration's forward and backward passes that compute the gradients. Therefore, an optimized solution could keep the auxiliary variables sharded across iterations until they are needed by checkpointing or summary. In practice, there are different patterns that could affect the placements of the all-gathers.

\begin{figure}[ht]
\centering
\includegraphics[width=0.45\textwidth]{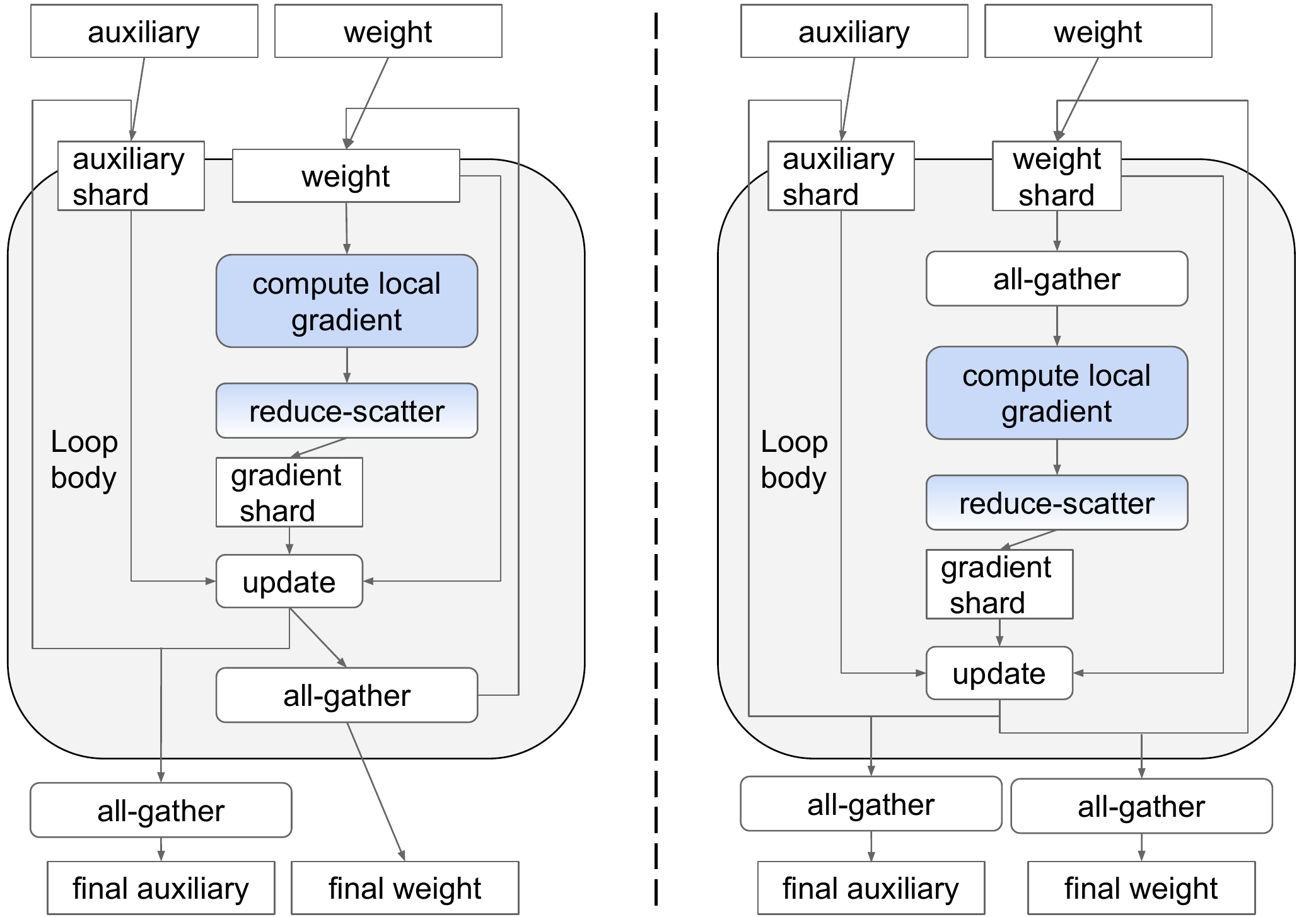}
\caption{\label{f:sharding-aux-loop} Two ways of sharding auxiliary variables with a loop. Left: only keep auxiliary sharded across iterations. Right: keep auxiliary and weight sharded across iterations, and all-gather the weight before forward/backward passes. Details will be discussed in Section~\ref{transform-on-loop}.}
\end{figure}

\begin{itemize}
    \item {\bf Compiler-visible loop.} If the compiler (graph optimization) could see a training loop in the graph, it could perform all-gather of auxiliary variables after the loop, amortizing the cost (see Figure~\ref{f:sharding-aux-loop}). If not, it would require additional help from the runtime system.
    \item {\bf Other uses of the auxiliary variables.} Although auxiliary variables are only used at weight-update time for the purpose of training, models in practice often include custom logic such as getting a summary of the current training progress which may use the complete state of variables. Such operations may be inside the training loop body, but often guarded by a conditional so that they only happen every ${k}$ steps.
\end{itemize}

We will discuss these issues in Section~\ref{analysis} and Section~\ref{transform}. The rest of the paper also addresses the following challenges that are critical to performance.

\begin{itemize}
    \item {\bf Sharding format.} How a tensor is divided across different replicas can be tricky on accelerators with tiled memory layouts~\cite{tpu-perf-guide}, since data formatting can be expensive. Additionally, individual dimensions on a tensor can be too small or not evenly-shardable among the replicas. To make the sharding of tensors efficient, our system chooses a set of cheap reformatting steps that could be efficiently fused into the sharding/unsharding operations.
    \item {\bf Non-elementwise optimizers.} With some model optimizers, the weight update computation may include non-elementwise operations. For example, some optimizers~\cite{adafactor,lars} use the weight norm or root-mean-square which involves \texttt{reduce} operators. We will discuss solutions of running non-elementwise computations on sharded data.
    \item {\bf Communication in large topology.} When the number of replicas is large, the shard size of a tensor can be very small such that the reduce-scatter and all-gather would become latency-bound. In such cases, our system will choose to partially shard the weight update computation among subgroups of replicas, and use batched communication operations to reduce the latency on large network topology.
\end{itemize}

%% file: analysis.tex
\section{Static Analysis}
\label{analysis}
XLA operators are relatively low-level compared to those in front-end frameworks like TensorFlow, which has two implications. First, lots of structural information about the model is lost when lowered to XLA (e.g., what parts of the graph are weight updates), which requires us to use static analysis to identify the operators to shard. Second, the set of operators is small, which makes analysis easier. We use static analysis to guarantee correctness and to identify transforms that are beneficial to performance.

\subsection{Correctness: cross-replica redundancy}
Weight update is a subset of the training graph that is redundant across replicas: since it does not have a batch dimension, all replicas are repeating the same computation on the same data. Redundancy is the property of an operator that it must output the same value in all replicas. Therefore, as long as an operator is redundant across replicas, it is safe to shard it across replicas.

\paragraph{Sources of redundancy.} There are three types of operators that are known to produce the same results, thus can be regarded as the sources of the analysis.

\begin{itemize}
    \item {\bf Constants.} Because all replicas are executing the same program, compile-time constants must be the same across replicas.
    \item {\bf Output of all-reduce.} By definition, an all-reduce operator produces the same output on the participating replicas. An exception is all-reduce operators with subgroups, where each group of replicas perform their own all-reduce, which could still be used in partial sharding within those subgroups, but we skip that case for simplicity.
    \item {\bf Annotated parameters.} The above obvious sources are insufficient to identify the weight update computation, because the initial values for the weight variables are passed in as parameters to the XLA graph, and XLA does not assume all replicas to have the same parameter values. Fortunately, in practical use cases of data-parallel training, these initial weight values are set to the same across replicas in order to keep the weights in sync. In our approach, the front-end framework, e.g., TensorFlow, needs to annotate the corresponding parameters in XLA to indicate that they will receive the same values during execution.
\end{itemize}

\paragraph{Propagation.} With the initial source set of redundant operators, we can run a propagation pass to identify other redundant operators. The analysis pass visits one operator at a time in topological order, i.e., producer before consumer.

For an operator that does not involve control flow, the analysis checks whether it has side effects or randomness, and whether all of its inputs are redundant. If all checks pass, this operator is marked as redundant.

Control flow in XLA is represented as special operators calling to nested computations.

\begin{itemize}
    \item {\bf A conditional} has a predicate and multiple branch computations. To determine a conditional's redundancy, the analysis first checks whether the predicate is redundant, then runs on all the branches to check their return values' redundancy. If all checks pass, the conditional can be marked as redundant.
    \item {\bf A while loop} has a body computation and a condition computation, which share the same input. The body's output is passed as the next iteration's input, so the output's redundancy must be used to determine the input's redundancy. To model this back edge, the analysis maintains tentative results of the operators in the loop, and runs iteratively on the condition and body, until a fixed point is reached. During each run, if the condition's result is determined to be non-redundant, all operators must be marked as non-redundant as well, since the control flow is different across replicas; this is unlikely to happen on the main training loop since all replicas are expected to execute the same number of steps.
\end{itemize}

\subsection{Performance: sharding profitability}
\label{perf-analysis}
\begin{figure}[ht]
\centering
\includegraphics[width=0.4\textwidth]{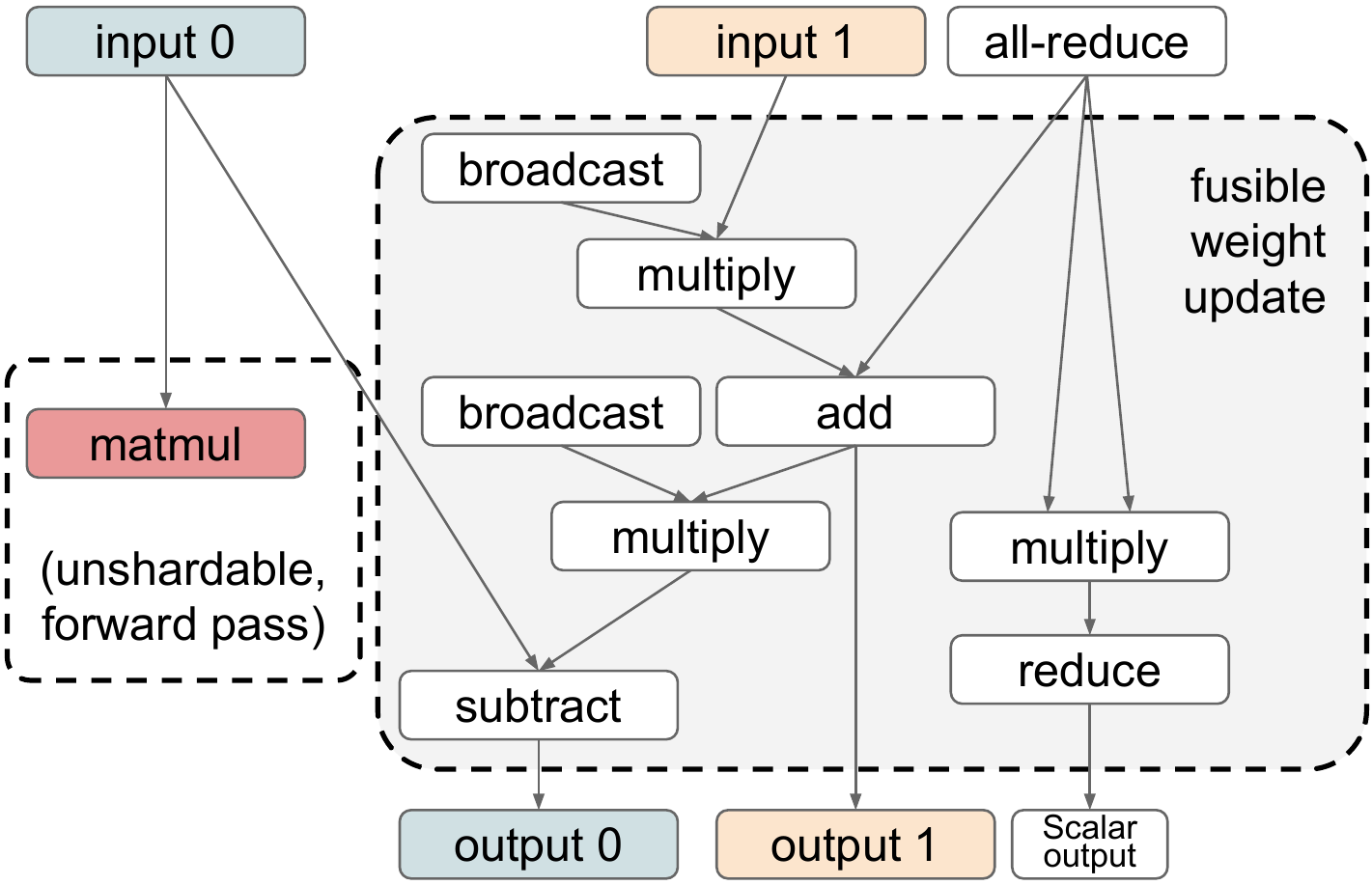}
\caption{\label{f:weight-update-around-ar} Weight update operators around all-reduce. If this is in a loop, Input 1 and Output 1 can be sharded across iterations, and there will be just one all-gather needed for Input 0 and Output 0 (either before Output 0 or before the matmul).}
\end{figure}

We analyze whether efficient sharding can be applied to each part of the identified redundant computation. Since an all-reduce operator precedes the update of a weight and the associated auxiliary variables, our analysis is centered around the all-reduce operators. The analysis first finds the cluster of redundant operators connected to each all-reduce, using simple propagation. Because weight updates are usually distinct parts of the training graph that do not have much interaction with the forward and backward passes, such propagation does not need to be overly sophisticated. Figure~\ref{f:weight-update-around-ar} shows a typical example of the identified clusters.

For each weight update, the impact on performance is primarily determined by two factors: the size reduction in local weight update and the requirement for communication. If the effect of size reduction overweighs the communication overhead, sharding can be applied. The calculation can be based on a fairly conservative cost models, since in typical cases (Figure~\ref{f:sharding-aux-loop}), sharding should give an obvious speedup.

Size reduction in local weight update must consider fusion; a good estimate is to use the combined size of the non-fusible inputs and outputs, instead of the number of operators sharded.

Because we always need only one reduce-scatter (Figure~\ref{f:sharding-aux-loop}), communication requirement is determined by the all-gathers needed for unshardable operators with sharded inputs. An unshardable operator can be part of the output of the program, a non-redundant operator, or an operator with unimplemented sharding transformation. There are also conditionally shardable operators, i.e., those supported only for certain sharding formats of the tensor. For example, a reduce operator along specific dimensions may not be supported for arbitrary reformatting. See Section~\ref{sharding-rep} for detailed discussion.

The placement of all-gather operators is heavily affected by control flow. In Figure~\ref{f:sharding-aux-loop}, only one all-gather is inside the loop, so that the amortized overhead of the extra all-gather operators is very small. The analysis accounts for this effect by marking the corresponding input/output pair of the loop as shardable, with the requirement that they must be sharded in the same way.

The training loop is critical to amortize all-gather cost for auxiliary variables. However, it is also common that there is not a compiler-visible loop, where the XLA graph only represents a single step, and the training loop could be entirely implemented by the user as a Python loop. We will discuss such cases in Section~\ref{transform-on-loop}.

We have seen models that transfer the on-device tensors to the host once in a certain number of steps, as a way to checkpoint, summarize, or debug the current training state. This is typically done using a conditional operator after the weight update, which contains an outfeed operator of the full weight and auxiliary tensors in one branch. For such cases, we have an analysis that estimates the frequency of different branches, and if the full tensors are only needed in an infrequent branch, we can place an all-gather inside that branch without adding much overhead. We implemented the frequency analysis by checking the conditional predicate's use of the loop induction variable, which is capable of recognizing the pattern described above.

%% file: transform.tex
\section{Graph Transformation}
\label{transform}

After the analysis passes, whether to shard each weight update is determined. This section discusses issues in implementing the sharded weight update efficiently, including how a weight tensor is sharded and the placement of all-gather operators in different scenarios. Performance of reduce-scatter and all-gather will be discussed later in Section~\ref{comm}.

\subsection{Sharding representation}
\label{sharding-rep}
For a set of weight-update operators (Figure~\ref{f:weight-update-around-ar}), all the inputs (gradients and the original weights and auxiliary variables) must be sharded in the same way, because they are consumed by the same set of elementwise operators during weight update. Without weight-update sharding, although all-reduce is also implemented as a reduce-scatter phase and an all-gather phase, it can choose arbitrary sharding internally because the sharding does not need to be exposed to other operators; in contrast, with sharded weight update, the sharding format used by the communication primitives must match the sharding on the inputs.

A weight tensor is represented as a multi-dimensional array. In processors like Cloud TPUs which have tiled memory layouts~\cite{tpu-perf-guide}, splitting some dimensions can be more expensive than splitting other dimensions. The chosen sharding must also be supported by the reduce-scatter and all-gather operators. Therefore, we always choose a dimension that is efficient for sharding and easier to be supported in reduce-scatter and all-gather.

\begin{figure}[ht]
\centering
\includegraphics[width=0.46\textwidth]{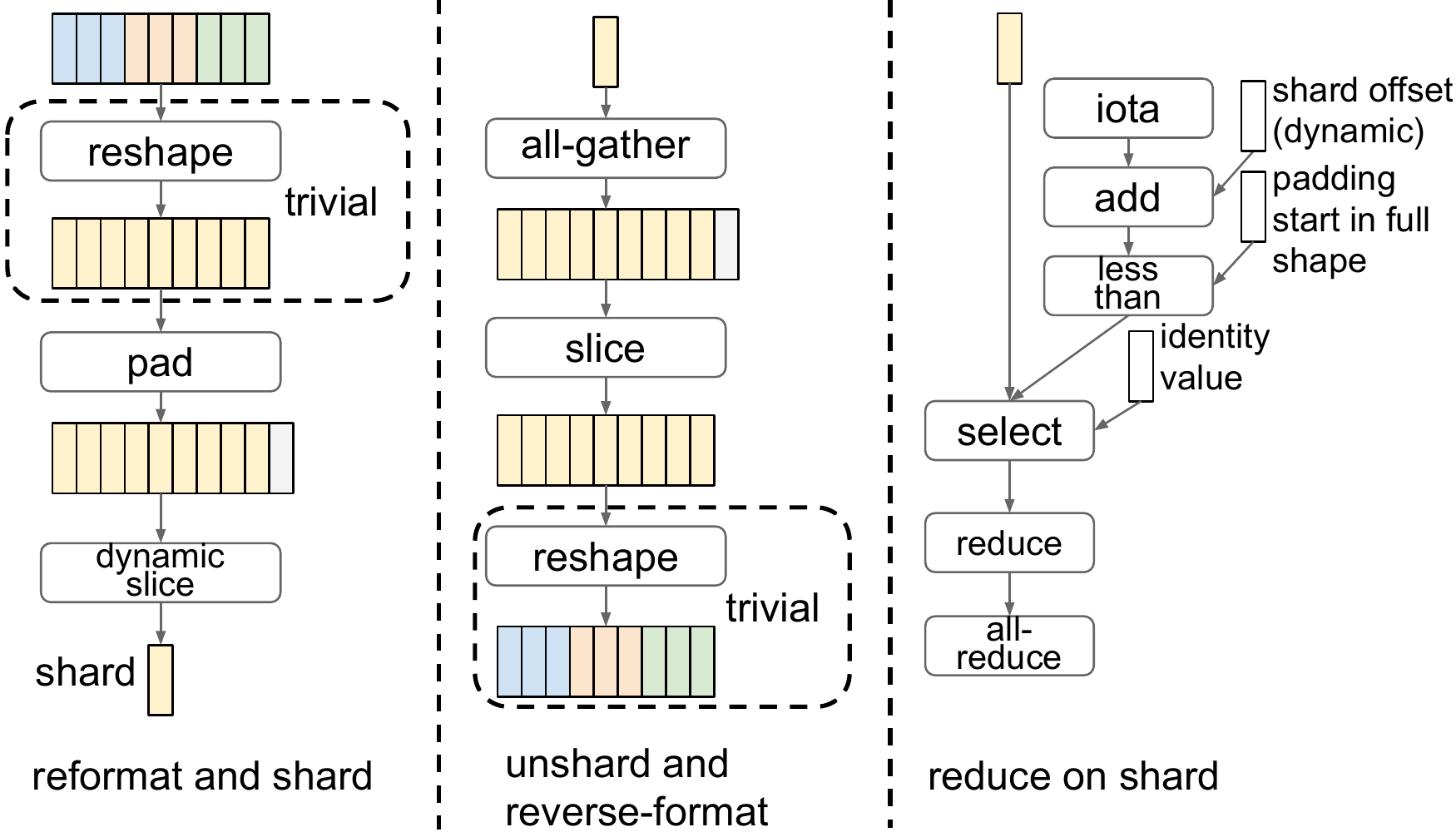}
\caption{\label{f:sharding-rep} Sharding and unsharding with reformatting. The right graph shows an example of handling non-elementwise operators on a shard with reformatting.}
\end{figure}

\paragraph{Data formatting.} One common problem is that the desired sharding dimensions are not evenly divisible by the number of shards (replicas). For example, ResNet~\cite{resnet} has weights with shape [3,3,256,256], where [3,3] are the desired sharding dimensions, but the shard count can be 8. To address such problems, we allow a tensor to be reformatted before sharded across replicas. Therefore, the sharding of a tensor is represented as a sequence of data formatting operators, followed by a dynamic-slice operator, as shown in Figure~\ref{f:sharding-rep}. The dynamic-slice specifies the dimensions to shard, and uses the replica-id to calculate the offset of the shard for each replica.

The formatting operators can include reshapes that combine dimensions, and pads that make the dimensions divisible by the shard count. Combining dimensions usually happens before padding, which helps minimizing the amount of padding. For example, [3,3,256,256] can be reshaped to [9,256,256], which allows it to be padded into [10,256,256] if the replica count is 10, instead of padding to [10,3,256,256] or [4,5,256,256]. The reformatting must be efficient to implement for the platform. In practice, we only choose reshapes that are trivial, which do not require any data movement. For Cloud TPUs, reshaping [3,3,256,256] into [9,256,256] can be trivial, but reshaping it to [589824] may be expensive due to the tiled memory layout.

There is another platform-dependent reformatting operator, bitcast. It means to reinterpret the on-device memory as a different shape, as long as the new shape's on-device representation does not go out of bound. Bitcast does not have consistent semantics in platform-agnostic XLA, but it is consistent for a specific platform, which is sufficient to guarantee that different inputs of the weight update are sharded in the same way. For example, for Cloud TPU, we can bitcast [3,3,256,256] into [576,8,128], making it shardable across 64 replicas without any padding.

In addition, we only choose reformatting operators that could be efficiently fused into operators around them. For example, the pad operator should be fused into the dynamic-slice so it does not access the entire memory buffer of the full shape.

\paragraph{Non-elementwise operators.} While most operators in weight update are simple elementwise arithmetic ones, some optimizers~\cite{lars,adafactor} also include non-elementwise operators, with the most common one being reduce.

Non-elementwise operators may impose restrictions on how a tensor can be reformatted. In a reduce operator, some dimensions are collapsed using the reduction function, and others are passed-through to the result; the reformatting is disallowed to combine a collapsed dimension with a pass-through dimension, through either reshape or bitcast. This does not restrict a reduce-to-scalar operator since all dimensions are collapsed.

Another restriction is for padding. Padded data elements in the collapsed dimensions could affect the result of the reduce, so they must be masked off with the identity value (e.g., 0 for addition and 1 for multiplication). This requires that the locations of the padding data must be identifiable after reformatting. If the source of padding is introduced as an explicit reformartting step without reshape or bitcast following it, the locations are identifiable as specified in the pad operator; instead, the restriction is often for implicit padding in bitcast: the tiled memory layout already implies padding, so reinterpreting a memory buffer could lose some of the padding information. Therefore, depending on the platform's memory layouts for tensors, certain bitcasts may introduce complexity when supporting reduce operators. The restriction depends on the implementation, and should avoid unsupported cases.

If sharding affects the collapsed dimensions, extra handling is required for the reduce operator. First, each replica needs to mask off the padded data. The padding areas on different replicas are different, depending on their shards' locations in the full shape, which requires the masking to be dynamic in the same training program. As shown in Figure~\ref{f:sharding-rep}, this can be achieved by comparing the elements' locations (iota + start offset) with the padding areas' locations on the full shape, then selecting between the shard data and the identity value based on the comparison results. Second, replicas need to combine their reduce results using an all-reduce. This is because the collapsed dimensions are lost in the reduce result, so they cannot be sharded, but each replica's local result is different from others and only captures data from its own input shard.

\subsection{Transform the training graph}
\label{transform-on-loop}
As discussed in Section~\ref{perf-analysis}, the placement of all-gather operators is critical to performance. With the help of the a training loop, we often need only one all-gather inside the loop.

\paragraph{Out-of-loop all-gather placement.} With a compiler-visible training loop, the all-gather operator for auxiliary variables can be placed after the loop, followed by required reverse formatting operators. Correspondingly, the original auxiliary variable values need to be sharded before the loop starts, using the reformatting operators and dynamic-slice as in Figure~\ref{f:sharding-rep}.

If there is not a compiler-visible loop, it is still possible to benefit from weight-update sharding by moving the sharding and unsharding of auxiliary variables outside of the training step program. One solution is to generate three separate programs after graph transformation: a sharding program, a main program, and an unsharding program. The sharding program contains the sharding operators of the variables before the training loop; the main program contains the training step with sharded weight update; the unsharding program contains the all-gather operators to reconstruct the full variables. It is the run-time system's responsibility to invoke each program at the right time. For instance, if the run-time system manages the training loop, it can invoke the sharding/unsharding programs before and after the loop; if even the run-time does not see a loop structure, it can still maintain states that track whether each variable is sharded, and conditionally invoke the sharding/unsharding programs when there is a state mismatch.

\paragraph{In-loop all-gather placemnt.} In Figure~\ref{f:sharding-aux-loop}, we have shown two potential ways to place the all-gather for the weight to be consumed by the forward and backward passes. The left graph shows the obvious way where the all-gather is at the end of the training step, and weight is already in full shape when the next iteration starts. The right graph instead keeps the weight sharded across loop iterations, like for auxiliary variables, but performs the all-gather right before it is needed by the forward and backward passes.

\begin{figure}[ht]
\centering
\includegraphics[width=0.38\textwidth]{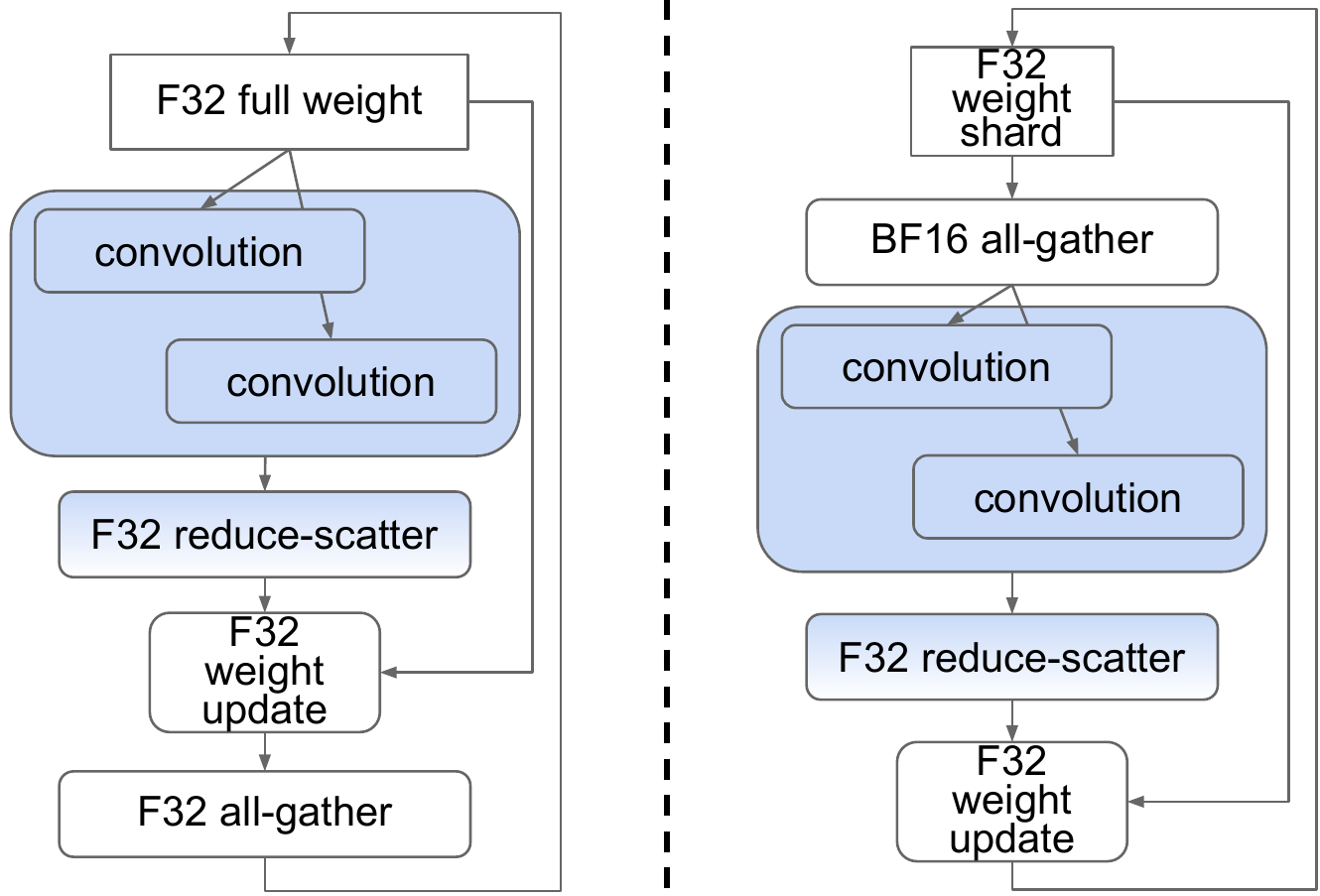}
\caption{\label{f:ag-bf16} By keeping the weight sharded across iterations (right graph), the full weight is only needed in bfloat16 precision.}
\end{figure}

\begin{figure*}[ht]
\centering
\includegraphics[width=0.75\textwidth]{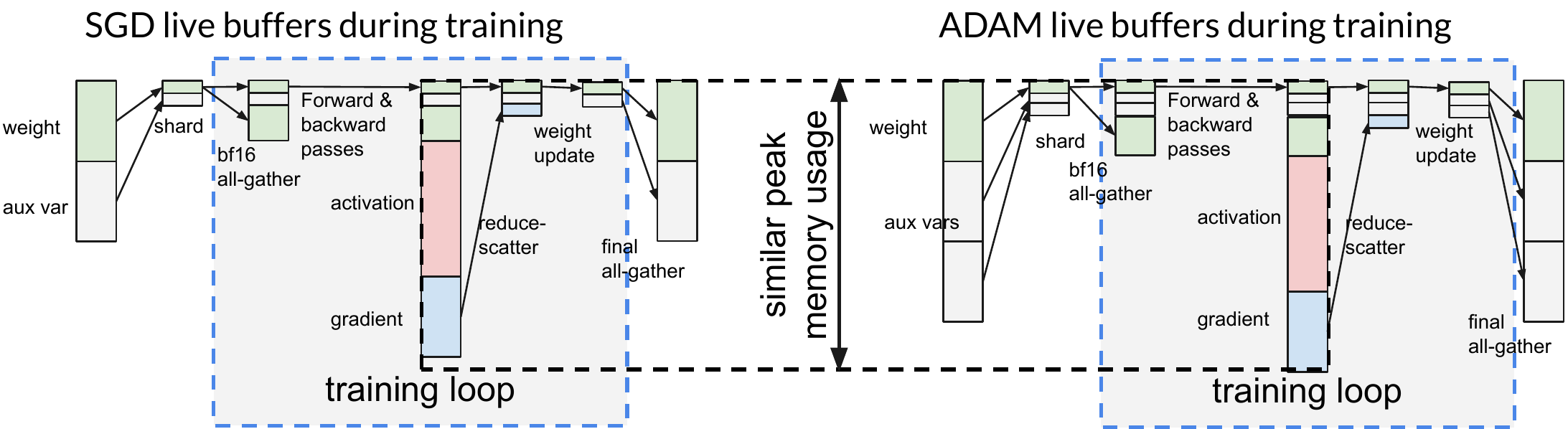}
\caption{\label{f:peak-memory} Reduced buffer live ranges of auxiliary variables allow ADAM to have similar peak memory as SGD.}
\end{figure*}

It may appear that the first approach is better for performance since it does not need the all-gather for the weight after the loop, even though that should be only a small amortized cost. However, we found in practice the second approach often enables more advanced optimizations. The main difference is that in the second approach the weight update no longer depends on the full weight. Weight update only requires the sharded data that is given when the step starts, and the full data after all-gather is only consumed by the forward and backward passes. In many image and language models, the forward and backward passes use the weight as an input to convolutions or matrix multiplies, which often have lower precision requirements on their inputs. For example, in typical training with Cloud TPUs, the precision of the input to a convolution is reduced to bfloat16~\cite{bfloat16}, while the weight update is often required to be in float32. With the second approach, the all-gather for the full weight can be performed in bfloat16 as shown in Figure~\ref{f:ag-bf16}, which dramatically reduces the amount of memory access and communication. This precision optimization is done automatically by a dataflow-based precision propagation pass.

\paragraph{Memory saving.} With the above transformation, the live ranges of weights and auxiliary variables are reduced. Especially for auxiliary variables, the full buffer is only required outside the training loop. Therefore, their buffers can be reused to store activations and gradients in the forward and backward passes. As shown in Figure~\ref{f:peak-memory}, this allows optimizers with different auxiliary variable sizes to have similar peak memory usage. More precisely, suppose the total size of weights is $W$, total size of auxiliary variables is $V$ (optimizer-specific), and the peak size of live activations and gradients in forward and backward passes is $P$, then our technique reduces peak memory usage from $W+V+P$ to $max(W+V/N+P, W+V)$ where $N$ is the number of shards. This allows the ADAM optimizer to be as efficient as SGD in terms of both performance and memory.

%% file: comm.tex
\section{Efficient Communication}
\label{comm}
Efficient reduce-scatter and all-gather implementation is important for performance even if the theoretical amount of communication is comparable to the all-reduce without weight-update sharding. There are two challenges, matching the sharding representation specified on the tensor (Section~\ref{sharding-rep}) and avoiding latency-bound communication on small shards.

\subsection{Fusion with data formatting}
\begin{figure}[ht]
\centering
\includegraphics[width=0.36\textwidth]{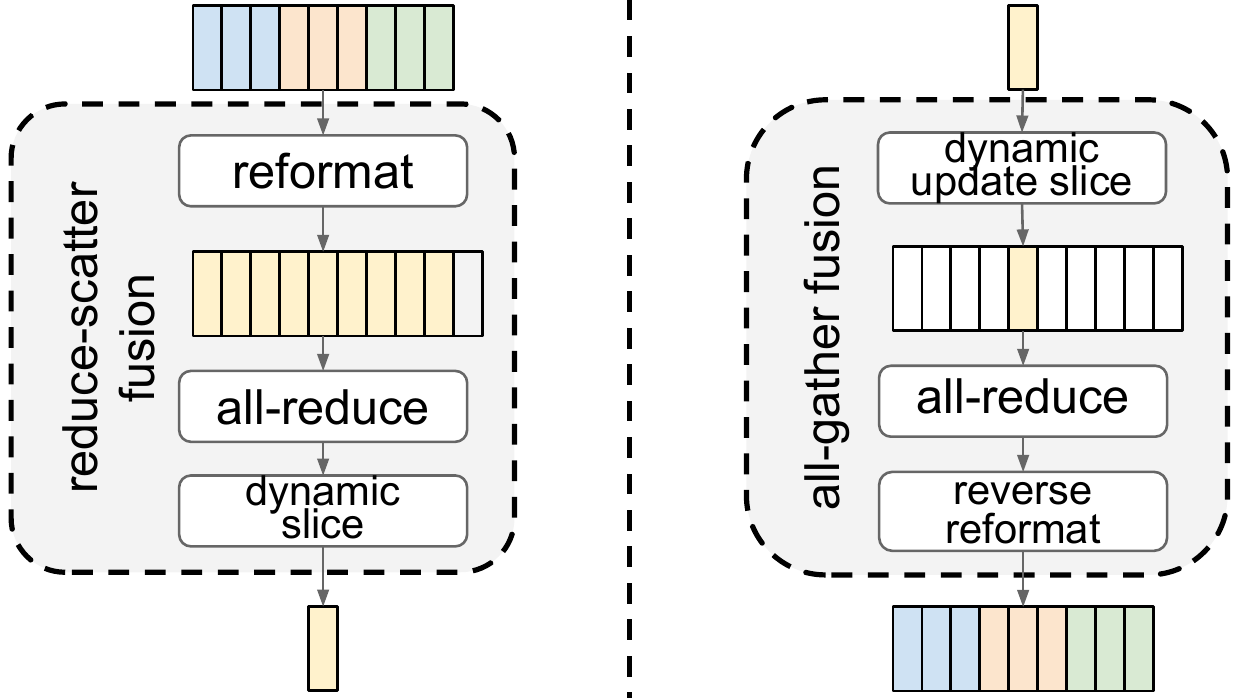}
\caption{\label{f:ar-fusion} Reduce-scatter and all-gather represented as fusion with reformatting and all-reduce.}
\end{figure}

The formatting steps chosen for each tensor in the sharding representation are needed to determine how it is divided into shards. If we pad the gradient before reduce-scatter, it would require each replica to perform local read and write on the full data. To avoid such inefficiency, we fuse the formatting operators into the reduce-scatter and all-gather. With the fusion representation, we can express flexible sharding without introducing complex configurations on the operators; in fact, we do not even need to define dedicated reduce-scatter or all-gather operators, because they can be expressed using all-reduce as shown in Figure~\ref{f:ar-fusion}.

In a classic algorithm of reduce-scatter and all-gather on $N$ replicas, the data is partitioned into $N$ pieces, and replicas form a logical ring and exchange pieces with neighbors in multiple rounds~\cite{thakur05}. In our fusion implementation, the boundaries of these pieces must exactly match the sharding format, and the padding is done in-place when preparing the data pieces.

The implementation of the fusion operators also guarantees that the shard assigned to a replica matches the location of it in the logical ring, so that the classic algorithm will produce the desired shard on each replica at the end. Because it is critical for the logical ring to utilize the bandwidth of the physical network's links, we choose shard ID based on the network topology, not the other way around. In practice, reduce-scatter and all-gather can be implemented in multiple phases in order to leverage specific network topology~\cite{cho19}. For instance, for an $N\times M$ array of devices, reduce-scatter with data size $D$ can be done first for each row with $D/M$ as the shard size, then for each column with $D/(MN)$ as the shard size. In such cases, the shard ID is calculated based on the topology of all phases.

\subsection{Utilizing network bandwidth for large topology}
In large-scale training where the number of replicas is large, the shard size of a weight or gradient tensor can be very small. For instance, a Cloud TPUv3 pod has 2048 cores (with 2 cores sharing a chip), so if a 4~MB tensor is partitioned in 2048 ways the shard size will be just 2~KB. First, an obvious problem is that the communication can easily become latency-bound; second, the small shard itself might require a significant amount of padding in a tiled memory layout, so that the effective transferred data size could be much larger than the full tensor.

\paragraph{Partial sharding.} In practice, sharding the weight update in 2048 ways does not provide observable saving compared to sharding in 64 ways, because the sharded weight-update time is already small compared to the rest of the training step. Therefore, we can choose to organize the replicas into independent  groups, and each group performs its own sharding. However, a per-group reduce-scatter only produces partial result since it does not accumulate the data from other groups. Therefore, an all-reduce across groups is needed after the reduce-scatter.

\begin{figure}[ht]
\centering
\includegraphics[width=0.46\textwidth]{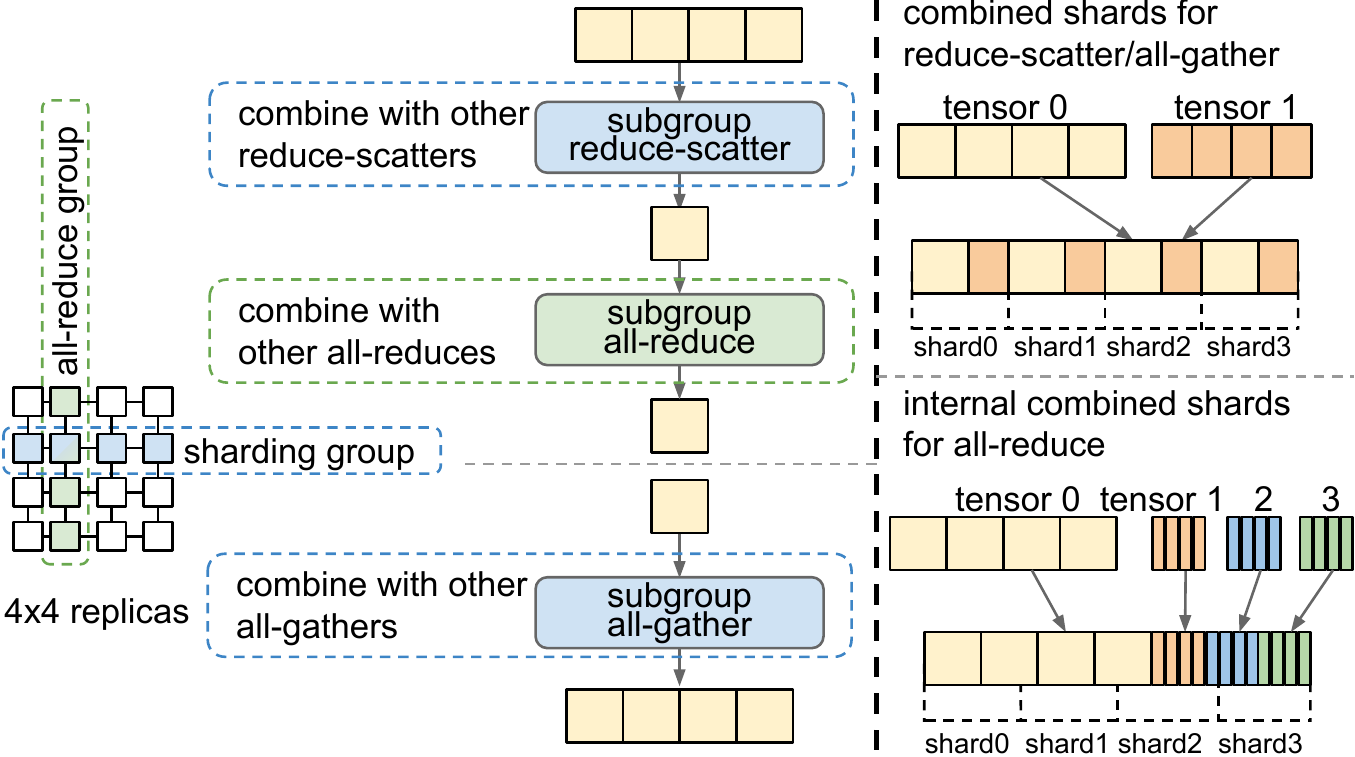}
\caption{\label{f:partial-sharding} Partial sharding and batched communication.}
\end{figure}

For an $N\times M$ array of replicas, the sharding groups can be defined as the $N$ rows, and the all-reduce will be performed on each of the $M$ columns (Figure~\ref{f:partial-sharding}). It may appear that the communication will still be latency-bound, because the all-reduce happens on the already sharded output of reduce-scatter, so the internal shard size of the all-reduce is still $D/(MN)$. In fact, as we show next, the grouping helps by enabling more aggressive batching of small data transfers.

\paragraph{Batched communication operators.} The weight update computations for different weight variables are typically independent from each other, so we can combine their communication operators together. This is possible because they share the same assignment of the groups and shards determined by the network topology.

A combined reduce-scatter or all-gather must maintain the original shard assignment for each tensor. To achieve this, each combined shard consists of one shard from every tensor. If there is excessive padding on one tensor's shards, it is likely to remain in the combined shard. Also, tracking these sharding boundaries is challenging in multi-phase reduce-scatter and all-gather.

By contrast, a combined all-reduce does not need to respect any sharding for inidividual input tensors, because its internal sharding does not need to be exposed. This makes implementing all-reduce on combined small tensors much more tractable and efficient. The input tensors can be conceptually concatenated together in full shapes, and the internal shards are partitions on the concatenated shape, as the right-hand-side graph shows in Figure~\ref{f:partial-sharding}. In addition to the all-reduce after a subgroup reduce-scatter, the all-reduce in Figure~\ref{f:sharding-rep} for the partial scalar reduce result can also be combined with other similar all-reduce operators.

As a result,  the partial sharding defers most of the handling of small shards into the combined all-reduce operators, where reduce-scatter and all-gather only perform combined operators in a single phase. This largely avoids the latency-bound communication for small shards. The batching of small communication operators is done automatically by a compiler pass.

%% file: eval.tex
\section{Evaluation}
\label{eval}
Automatic weight-update sharding is a key technique that enabled the state-of-the art training performance in Google's MLPerf-0.6 submission~\cite{mlperf0.6, google-mlperf0.6}.
We evaluated the performance improvements of several models with automatic weight-update sharding enabled. The models include ResNet-50~\cite{resnet}, Transformer~\cite{transformer} and NCF~\cite{ncf}. ResNet-50 and Transformer are based on the same configuration as in the MLPerf 0.6 submission. The test platform is Cloud TPUv3~\cite{tpu} with different topology configurations: 16 and 1024 chips in a 2D mesh, where each chip contains two processing cores. We use data-parallel training for all models, where each replica occupies a single core.

\begin{table}[ht]
    \centering
    \begin{tabular}{c|c|c|c}\toprule
        \textbf{Model} & \textbf{Core count} & \textbf{Batch size} & \textbf{Optimizer} \\\hline
         \multirow{2}{*}{ResNet-50} & 32 & 4096 & LARS \\\cline{2-4} 
          & 2048 & 32768 & LARS \\\hline 
         \multirow{2}{*}{Transformer} & 32 & 512 & ADAM \\\cline{2-4} 
          & 2048 & 2048 & ADAM \\\hline 
         NCF & 32 & 98304 & ADAM \\\bottomrule
    \end{tabular}
    \caption{Optimizer and batch size of evaluated models}
    \label{t:eval_configs}
\end{table}

\subsection{Performance}
\begin{figure}[ht]
\centering
\includegraphics[width=0.4\textwidth]{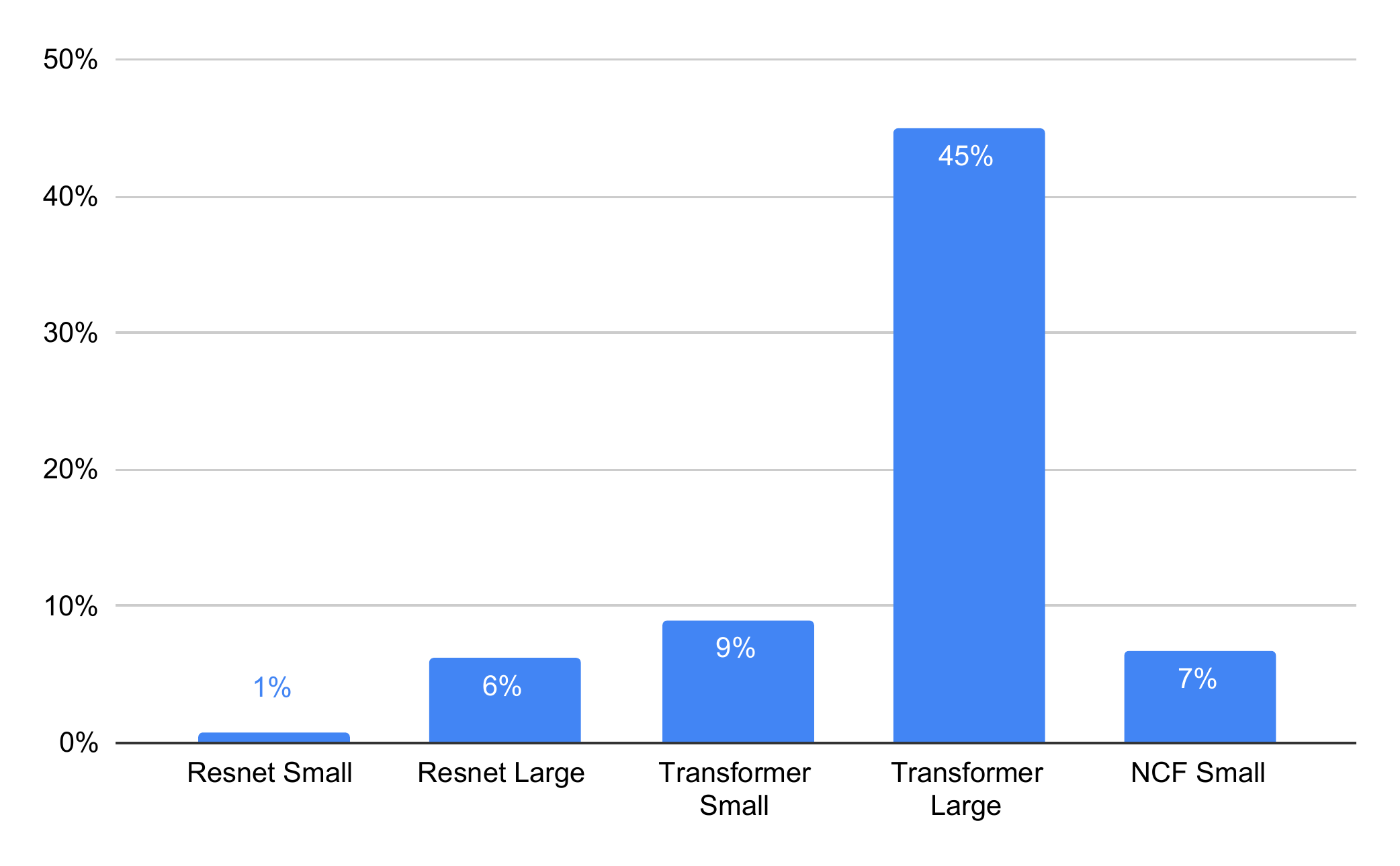}
\caption{\label{f:perf_eval} Step time reduction when automatic weight-update sharding is enabled.}
\end{figure}
Figure~\ref{f:perf_eval} shows the performance improvements of automatic weight-update sharding against the replicated weight update for different models.

At small scale (16 TPUv3 chips), we keep the per-replica batch size as large as possible to maximize TPU utilization. In this setup, the step time is relatively long and the constant weight-update time is amortized. As a result, for models with small weight sizes, the performance impact is small. But we still observes improvements in the range of 9\% for language models like Transformer where weight sizes are large.

At large scale (1024 TPUv3 chips), we decrease the per-replica batch size to keep the global batch size reasonably small. As a result, the step time reduces and the performance impact of automatic weight-update sharding becomes much larger. Even for image models like ResNet where weight sizes are small, it is giving a 6\% speedup. For Transformer, the step time reduces from 46.5ms to 25.6ms when we enable this optimization.

\subsection{Memory saving}
\begin{figure}[ht]
\centering
\includegraphics[width=0.4\textwidth]{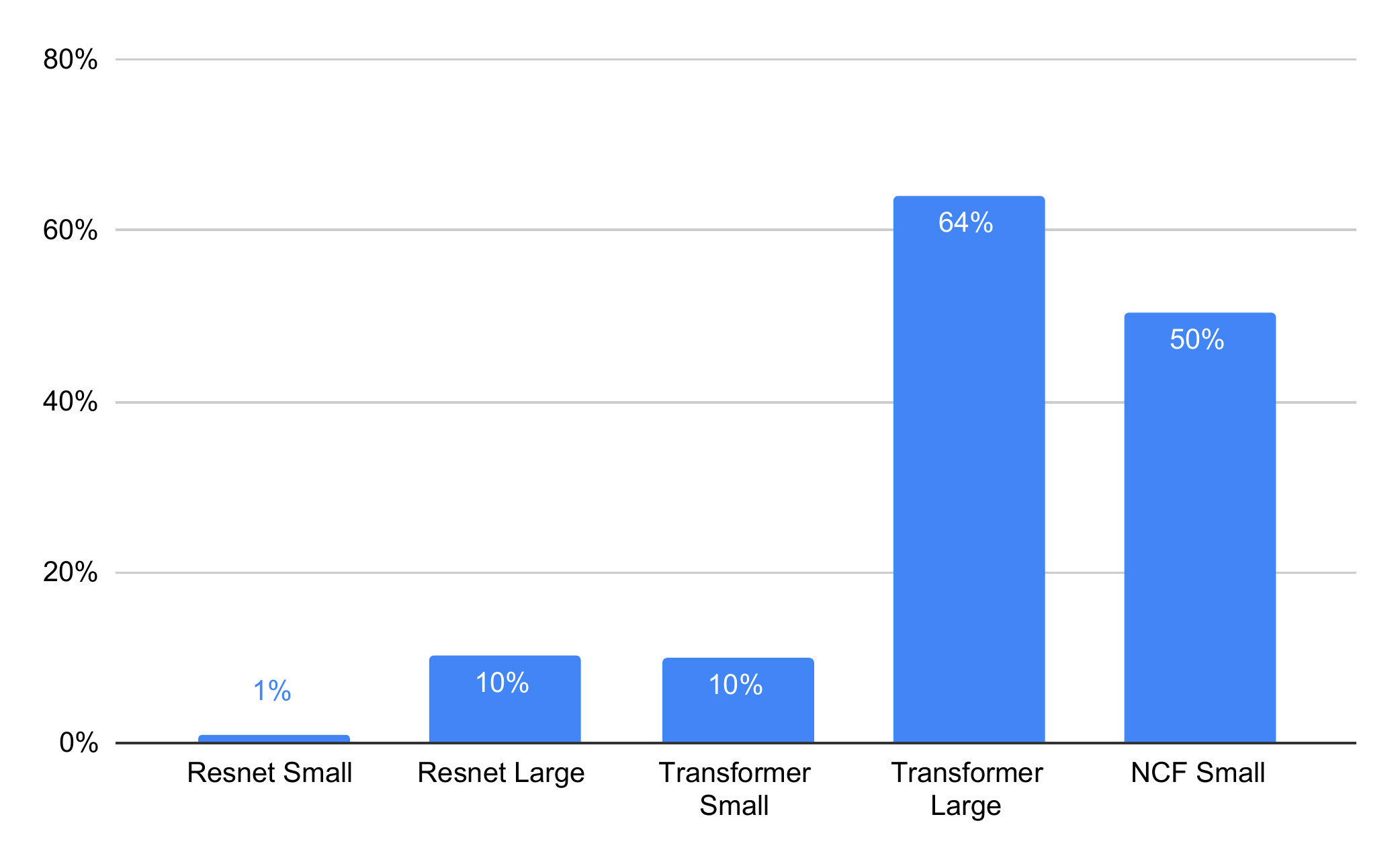}
\caption{\label{f:mem_eval} Activation memory saving when automatic weight-update sharding is enabled.}
\end{figure}
With auxiliary variables sharded during the training steps \ref{f:sharding-aux-loop}, their live ranges are split into two small segments before and after the training loop. Therefore, in the training loop body, their buffers could be reused by activations or intermediate results, which reduces peak memory usage. The actual saving is determined by the memory allocator, which is subject to problems like fragmentation.

Figure~\ref{f:mem_eval} shows the activation memory saving from this optimization. For models like Resnet, where weights are small comparing with activations, and there is only one copy of full shape auxiliary variable from SGD, the reusable memory is small. For models like NCF where activation size is comparable to weight size, and optimizer (e.g. Adam) creates two copies of the full shape auxiliary variables, there are more memory to reuse thus savings are larger.

%% file: related.tex
\section{Related Work}
\label{related}

\paragraph{Partitioned parameter servers.} In asynchronous training, parameter servers \cite{distblief,li14-param-server} are often used for weight update, where partitions of weights can be sharded across different server instances. The major difference from our approach is that these systems partition weights across multiple server instances, while our approach shards weight update across the existing workers (replicas) without using extra resources. Also, asynchronous training is a very different setting than our focus.

\paragraph{Parallel programming frameworks.}
Mesh-TensorFlow \cite{mtf} is a Single-Program-Multiple-Data (SPMD) framework that allows users to write programs with different tensor dimension split across dimensions of a multi-dimensional processor mesh. While weight update can be sharded using Mesh-TensorFlow, it is orthogonal to our work, because it requires each mesh dimension to have a specific meaning and if it is assigned to the batch (replica) dimension, it cannot be used to split weight update. In our work, the weight-update is split across replicas without using more processors than pure data parallelism. GPipe~\cite{gpipe} is a library for implementing pipeline parallelism for sequence models, which is orthogonal to our approach since it does not parallelize across replicas.

\paragraph{Automated parallelism.}
FlexFlow~\cite{jia19} uses automated search to discover the optimal partition of operations in a graph, which has a flexible search space to cover model and data parallelism. While it focuses on determining the partition strategy for every operation, our system also leverages global graph transformation for the training loop in order to amortize all-gather cost.

\paragraph{Manually designed parallelism.}
There are custom parallel training strategies designed for specific models \cite{gnmt,one-weird-trick}, which typically mix data and model parallelism. In contrast, our approach focus on the general weight-update pattern in synchronous data-parallel training, where the benefit varies across use cases.

%% file: conclusion.tex
\section{Conclusion}
\label{conclusion}
This paper presents a set of analyses and transformations for data-parallel deep learning training, which reduces the weight-update time by sharding across replicas. To minimize overhead of sharding, the approach carefully chooses communication patterns and data formatting, based on the training loop structure and the network topology of devices. It achieves significant speedups on language and large-scale image models, without requiring any additional devices.

%% file: main.bbl
\begin{thebibliography}{10}

\bibitem{mlperf0.6}
{MLPerf Training v0.6 Results}.
\newblock \url{https://mlperf.org/training-results-0-6}, 2019.
\newblock Online; accessed 15 August 2019.

\bibitem{xla}
{XLA: Optimizing Compiler for TensorFlow}.
\newblock \url{https://www.tensorflow.org/xla}, 2019.
\newblock Online; accessed 15 August 2019.

\bibitem{tensorflow}
{\sc Abadi, M., Barham, P., Chen, J., Chen, Z., Davis, A., Dean, J., Devin, M.,
  Ghemawat, S., Irving, G., Isard, M., Kudlur, M., Levenberg, J., Monga, R.,
  Moore, S., Murray, D.~G., Steiner, B., Tucker, P., Vasudevan, V., Warden, P.,
  Wicke, M., Yu, Y., and Zheng, X.}
\newblock {TensorFlow: A System for Large-Scale Machine Learning}.
\newblock In {\em {12th USENIX Symposium on Operating Systems Design and
  Implementation (OSDI)}\/} (Savannah, GA, Nov. 2016).

\bibitem{cho19}
{\sc Cho, M., Finkler, U., and Kung, D.}
\newblock {BlueConnect: Decomposing All-Reduce for Deep Learning on
  Heterogeneous Network Hierarchy}.
\newblock In {\em Proceedings of the Conference on Systems and Machine Learning
  (SysML)\/} (Palo Alto, CA, 2019).

\bibitem{distblief}
{\sc Dean, J., Corrado, G., Monga, R., Chen, K., Devin, M., Mao, M., aurelio
  Ranzato, M., Senior, A., Tucker, P., Yang, K., Le, Q.~V., and Ng, A.~Y.}
\newblock {Large Scale Distributed Deep Networks}.
\newblock In {\em Proceedings of the 26th Conference on Neural Information
  Processing Systems (NIPS)\/} (Lake Tahoe, Nevada, Dec. 2012).

\bibitem{tpu}
{\sc {Google Cloud}}.
\newblock {Cloud TPU}.
\newblock \url{https://cloud.google.com/tpu/}.
\newblock Online; accessed 15 August 2019.

\bibitem{tpu-perf-guide}
{\sc {Google Cloud}}.
\newblock {Cloud TPU Performance Guide}.
\newblock \url{https://cloud.google.com/tpu/docs/performance-guide}.
\newblock Online; accessed 15 August 2019.

\bibitem{bfloat16}
{\sc {Google Cloud}}.
\newblock {Using bfloat16 with TensorFlow models}.
\newblock \url{https://cloud.google.com/tpu/docs/bfloat16}, 2019.
\newblock Online; accessed 15 August 2019.

\bibitem{resnet}
{\sc He, K., Zhang, X., Ren, S., and Sun, J.}
\newblock {Deep Residual Learning for Image Recognition}.
\newblock In {\em Proceedings of IEEE Conference on Computer Vision and Pattern
  Recognition (CVPR)\/} (Las Vegas, Nevada, June 2016).

\bibitem{ncf}
{\sc He, X., Liao, L., Zhang, H., Nie, L., Hu, X., and Chua, T.}
\newblock Neural collaborative filtering.
\newblock {\em CoRR abs/1708.05031\/} (2017).

\bibitem{gpipe}
{\sc Huang, Y., Cheng, Y., Chen, D., Lee, H., Ngiam, J., Le, Q.~V., and Chen,
  Z.}
\newblock Gpipe: Efficient training of giant neural networks using pipeline
  parallelism.
\newblock {\em CoRR abs/1811.06965\/} (2018).

\bibitem{jia19}
{\sc Jia, Z., Zaharia, M., and Aiken, A.}
\newblock {Beyond Data and Model Parallelism for Deep Neural Networks}.
\newblock In {\em Proceedings of the Conference on Systems and Machine Learning
  (SysML)\/} (Palo Alto, CA, 2019).

\bibitem{adam}
{\sc Kingma, D.~P., and Ba, J.~L.}
\newblock {Adam: a Method for Stochastic Optimization}.
\newblock In {\em {International Conference on Learning Representations
  (ICLR)}\/} (San Diego, CA, May 2015).

\bibitem{one-weird-trick}
{\sc Krizhevsky, A.}
\newblock One weird trick for parallelizing convolutional neural networks.
\newblock {\em CoRR abs/1404.5997\/} (2014).

\bibitem{krizhevsky12}
{\sc Krizhevsky, A., Sutskever, I., and Hinton, G.~E.}
\newblock {ImageNet Classification with Deep Convolutional Neural Networks}.
\newblock In {\em Proceedings of the 26th Conference on Neural Information
  Processing Systems (NIPS)\/} (Lake Tahoe, Nevada, 2012).

\bibitem{google-mlperf0.6}
{\sc Kumar, S., Bitorff, V., Chen, D., Chou, C., Hechtman, B., Lee, H., Kumar,
  N., Mattson, P., Wang, S., Wang, T., and Zhou, Y. X.~Z.}
\newblock {Scale MLPerf-0.6 models on Google TPU-v3 Pods}.
\newblock {\em CoRR abs/1909.09756\/} (2019).

\bibitem{li14-param-server}
{\sc Li, M., Andersen, D.~G., Park, J.~W., Smola, A.~J., Ahmed, A., Josifovski,
  V., Long, J., Shekita, E.~J., and Su, B.-Y.}
\newblock Scaling distributed machine learning with the parameter server.
\newblock In {\em 11th {USENIX} Symposium on Operating Systems Design and
  Implementation ({OSDI} 14)\/} (Broomfield, CO, Oct. 2014), {USENIX}
  Association, pp.~583--598.

\bibitem{MPI-2.2}
{\sc {MPI Forum}}.
\newblock {MPI: A Message-Passing Interface Standard. Version 2.2}, September
  4th 2009.
\newblock available at: \url{http://www.mpi-forum.org} (Dec. 2009).

\bibitem{momentum}
{\sc Qian, N.}
\newblock {On the momentum term in gradient descent learning algorithms}.
\newblock {\em {Neural Networks} 12}, 1 (1999), 145 -- 151.

\bibitem{mtf}
{\sc Shazeer, N., Cheng, Y., Parmar, N., Tran, D., Vaswani, A., Koanantakool,
  P., Hawkins, P., Lee, H., Hong, M., Young, C., Sepassi, R., and Hechtman, B.}
\newblock {Mesh-Tensorflow: Deep Learning for Supercomputers}.
\newblock In {\em Proceedings of the 32nd Conference on Neural Information
  Processing Systems (NeurIPS)\/} (Montr\'eal, Canada, 2018).

\bibitem{adafactor}
{\sc Shazeer, N., and Stern, M.}
\newblock {Adafactor: Adaptive Learning Rates with Sublinear Memory Cost}.
\newblock {\em CoRR abs/1804.04235\/} (2018).

\bibitem{thakur05}
{\sc Thakur, R., Rabenseifner, R., and Gropp, W.}
\newblock {Optimization of Collective Communication Operations in MPICH}.
\newblock {\em Int. J. High Perform. Comput. Appl. 19}, 1 (Feb. 2005), 49--66.

\bibitem{transformer}
{\sc Vaswani, A., Shazeer, N., Parmar, N., Uszkoreit, J., Jones, L., Gomez,
  A.~N., Kaiser, L., and Polosukhin, I.}
\newblock {Attention Is All You Need}.
\newblock In {\em Proceedings of the 31st Conference on Neural Information
  Processing Systems (NIPS)\/} (Long Beach, CA, 2017).

\bibitem{gnmt}
{\sc Wu, Y., Schuster, M., Chen, Z., Le, Q.~V., Norouzi, M., Macherey, W.,
  Krikun, M., Cao, Y., Gao, Q., Macherey, K., Klingner, J., Shah, A., Johnson,
  M., Liu, X., Łukasz Kaiser, Gouws, S., Kato, Y., Kudo, T., Kazawa, H.,
  Stevens, K., Kurian, G., Patil, N., Wang, W., Young, C., Smith, J., Riesa,
  J., Rudnick, A., Vinyals, O., Corrado, G., Hughes, M., and Dean, J.}
\newblock Google's neural machine translation system: Bridging the gap between
  human and machine translation.
\newblock {\em CoRR abs/1609.08144\/} (2016).

\bibitem{lars}
{\sc You, Y., Gitman, I., and Ginsburg, B.}
\newblock {Scaling SGD Batch Size to 32K for ImageNet Training}.
\newblock {\em CoRR abs/1708.03888\/} (2017).

\end{thebibliography}
